\documentclass[reprint,amsmath,amssymb,prb]{revtex4-1}

\usepackage{bbding}
\usepackage[utf8]{inputenc}
\usepackage[T1]{fontenc}
\usepackage{graphicx}
\usepackage{dcolumn}
\usepackage{bm}
\usepackage{multirow}
\usepackage{subfigure}
\usepackage{hyperref}
\usepackage[american]{circuitikz}
\usepackage{tikz}

\DeclareMathAlphabet{\mathpzc}{OT1}{pzc}{m}{it}

\begin{document}

\title{Simulation and Experimental Verification of DNA Damage Due to X-Rays Interaction \vspace*{0.5em}}
\author{Alexandra Pinto Castellanos}
\address{Zurich, Switzerland}
\email{apintoca@student.ethz.ch}

\date{September, 2013}
\begin{abstract}
This paper describes the model for DNA in MATLAB taking into account all of component atoms. In this model, it is possible to generate sequences with length $10000$ basis pairs available for introducing all types of sequences. Once the strands are generated, it is studied the DNA damage in the single strand and double strand. The damage are outcomes of ionising radiation of X-rays when interacting with the DNA immersed in water. This is a theoretical and experimental in-vitro study that quantifies the single strand and double strand damage for different doses of radiation. This can be useful to predict the exact risks of expositions to radiations. In simulations, it is taken into account the damage caused by free electrons generated by the effect of the interaction with the water molecules, this is different to the effect considered in radiobiology, where indirect damages are due to chemical reactions. The spatial distribution of the electrons is obtained from Geant4 and here this distribution is used for the creation of rays as three-dimensional random trajectories through Monte Carlo simulations. It is also presented the experimental DNA damage through radiating DNA samples immersed in water with a X-rays unit with Molybdenum target. The damage level is quantified through Atomic Force Microscopy (AFM). It is possible to conclude a direct relation between the damage and the radiation doses with the experimental results.
\end{abstract}

\maketitle

\section{Introduction}
When cells are exposed to ionising radiation, the DNA molecules are damaged and this can lead to cell death \cite{sachs_dna_1992}. From the bandwidth of ionising radiation, X-rays is the most important radiation when interacting with the nuclear DNA in a cell target \cite{Mori2018}. X-rays are photons with frequencies from $50$ to $5000$ times higher than visible light and with wavelength in the order of an atom size, which make them relevant for its power at interacting with matter. X-rays are able to pass across most solids \cite{fasrt_principles_2009}.The electromagnetic spectrum is characterised by the wave-particle duality however, it is observed their behavior differently depending on their energy. Opposite to high energetic photons like X-rays, are radio photons which its detect in their wave form and high energetic photons like X-rays are detected more in their particle state. X-rays are created by emission of photons at electronic transitions due to acceleration and deceleration of charged particles \cite{attia_measurement_2011}.\\ \\
Studying of radiation is biologically important due to the effects of the particle interactions. Two factors are involve in cell damage after radiation: 1. Photon energy, 2. Dose rate \cite{10.1093/jrr/rrt222}. Normally cells have between 10 thousand to one million molecular lesions per cell per day due to environmental and metabolic process. For example, the Ultraviolet components of sunlight can cause more than $1\times 10^5$ lesions in the DNA molecules of one cell per day \cite{lord_dna_2012}. These lesions are repaired by the DNA reparation mechanism. However, the rate of this mechanism and its efficiency depend on the cell type, cell age, and extra-cellular environment \cite{panasci_dna_2010}. Hence, this mechanism is different for each person in terms of time and efficiency. Therefore, some people have a large amount of DNA damage and their cells cannot be repaired correctly which causes mutations \cite{popanda_radiation-induced_2003}.\\ \\
The Single-Strand Breaks, i.e., SSBs and the Double-Strand Breaks, i.e., DSBs \cite{lord_dna_2012} are the main damage of DNA molecules which are exposed to the X-ray radiations. Quantifying damage of DNA after X-rays exposition is crucial for radiotherapy and aerospace applications. The main contributions of this research are summarised in the following. I quantify the damage in the DNA molecules due to the interaction with X-rays and the free radicals generated in water. The develop of these studies is done by simulation of DNA interacting with the X-rays, obtaining the damage statistics according to the reactions between the DNA components, photons and secondary electrons generated by cascade effects \cite{friedland_track_2011}.\\ \\
When X-ray photons interact with the DNA immersed in water, they deposit their energy along the electron trajectories which are generated during the radiation process. They deposit $95\%$ of their energy through the Compton Effect, vibrations, ionizations and excitations. Photons interact directly with the atom's electrons and transfer part of their energy \cite{bernhardt_modeling_2003}. However, another important component of the radiation-DNA interaction is that photons also have an influence in the atoms's protons that compose the DNA molecules.\\ \\
In addition to the direct effect of X-rays on DNA molecules, it is necessary to consider the effect of X-rays on the surrounding molecules. DNA molecules are always surrounded by more molecules, mainly water, and these also interact with the X-rays. Hence, these processes generate free radicals which are harmful for the integrity of DNA molecules \cite{elshaikh_advances_2006}. To study the radiation effects in alive systems, it is required theoretical models for risk estimation and experimental studies for verification, those studies are out of the scope of this paper. Many researchers have investigated the numerical models however, the computational power is an important limiting factor for the risk estimation in different radiation doses. A drawback of other studies is the lack of correlation between the experimental results and quantifications are done by simulations and mathematical models \cite{deisboeck_multiscale_2011}.\\ \\
Track structure simulation for X-ray irradiation are commonly done in plasmid DNA when immerse in different chemicals by assuming distributions of Single Strand Break and Double Strand Break with respect to doses of radiation which is also assumed as a distribution. The secondary effects given by free radicals is usually computed by the mean free drift path of the radicals in the medium, most of the time water \cite{Brons2003}. In this respect, the software presented in this paper is more accurate with respect to the structure of the DNA, considering the position of its atoms. The damage is not assumed but computed after counting of interactions between DNA-atom to photons or radicals. The radiation distribution is not assumed but generated by Monte Carlo. A good introduction to track simulations and its stochastic considerations can be found at Curtis paper developed at NASA \cite{Curtis}\\ \\
Another important simulation component is the generation of the DNA, current DNA models have been developed under forceful geometrical parameterisation without considering the true atomic double-stranded structure of DNA molecules. An example is the PARTRAC simulation, which main focus is the physical phenomena of the interactions. The fact that the atomic spatial distribution is not simulated, results on the lack of considerations of the strong dependency of the DNA atomic components spatial distribution and the interactions with the embedded medium, approximately mainly water. \\ \\
One of the main achievement of this research is considering the atomic components of DNA molecules in the simulations, which creates more accurate results of the damage. Moreover, I assume DNA molecules immersed in water to take into account the generation of electrons released by water which are responsible of the main DNA damage. Another detail that this simulation considers is that both the DNA and water molecules are exposed to X-rays and the damage is quantified according to the number of DSB obtained by the DNA strand rupture and the total number of electrons that interact.\\ \\
The analysis of the results require the introduction of the formal unit used to quantify the radiation dose are Grays \cite{fasrt_principles_2009}. The rest of this paper is organised as follows. In Section II, it is expressed the differences between the simulations and two existing ones. In Section III, it is presented the detailed methodology to analyse the effect of X-rays on the DNA molecules. In Section IV, it is shown the simulation and experimental results. In Section V, it is discussed about the obtained results. Finally, I give concluding remarks in Section VI.

\section{Differences between Proposed Simulations and other codes}
\subsection{Other codes}
Let me compare the proposed simulation with two existing ones, these are not the only available software but they are the pioneers in the study of interactions between X-rays and DNA molecules. The first one, i.e., PARTRAC is a code which is developed during the last $25$ years by the Helmholtz Zentrum Munchen ISS, Research Group Radiation Risk \cite{friedland_track_2011}. This code consists of different modules whose interactions are described individually. These modules were programmed in FORTRAN. Fig. \ref{2} shows the modules of PARTRAC code.
\begin{figure}
\centering
\includegraphics[width=0.4\textwidth, height=0.2\textheight]{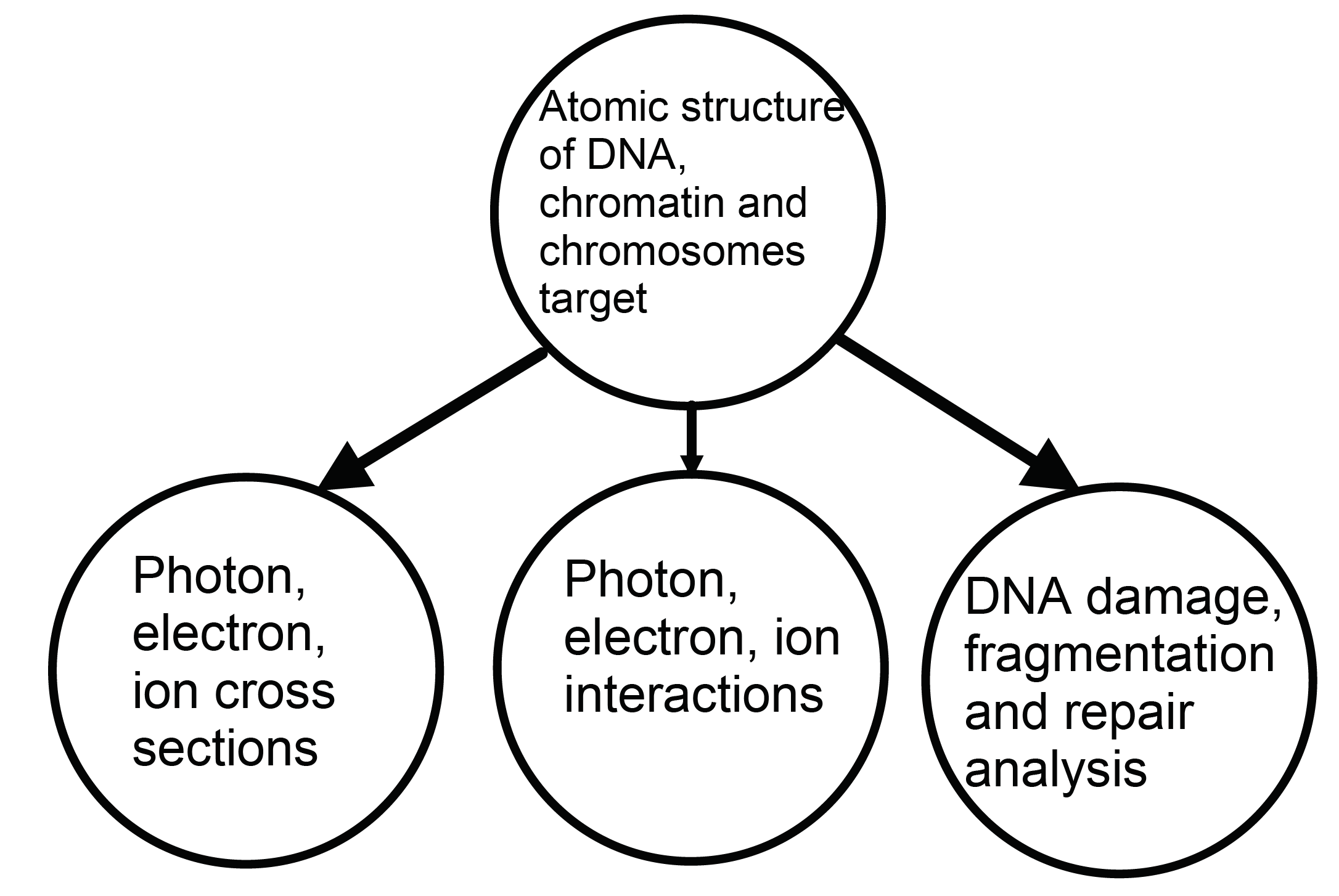}
\caption{Diagram of the modules that take part of the PARTRAC code \cite{friedland_track_2011}.}
\label{2}
\end{figure}
This code has the DNA organisation in their higher domains, i.e., modelling chromatin fibers and its packaged into chromosomes. In this DNA model, the double helix is assumed as a cylinder with a diameter equal to $2.3 \, \text{nm}$ which is fractioned into sections whose thickness is $0.34 \, \text{nm}$. These fractions represent the nucleotides. Moreover, developers have divided this cylinder into a smaller cylinder whose diameter is equal to 1 nm and two arcs rotated $36^{\circ}$ surround it. This is done for all nucleotides to achieve the basis separation of the sugar and the phosphate group \cite{friedland_track_2011}.
This parameterisation does not consider the DNA atomic composition which plays an important role in the interactions between electrons and DNA molecules. In this code, damage occurs according to the energy deposition in spherical volumes or cylindrical in nano metric scale; however, the atomic bonds are ignored while they are the cause of the majority of damage.
Moreover, this is a private software which makes its development slower since only a little group of researchers are allowed to use it. However, this software has the highest achievements in this field and it can even consider the DNA reparation process.

The second one, i.e., GEANT4 is a Software available in Linux Scientific which simulates interactions between particles and their trajectories effects along materials with variable characteristics. It mainly focuses on particles with high energy, however recent developments in Geant4-DNA, (as part of the European Spatial Agency (ESA) collaboration), is designed for low energy (~eV) simulations. Geant4-DNA consists of Monte Carlo simulations for crossing particles with lower energy through matter \cite{francis_molecular_2011}.
Its main objective is estimation of the risk of cancer for humans after the exposition. It employs a Monte Carlo simulation for generating X-rays and analyses different phenomena due to their interactions with water. In 2007, Ziad Francis, through interchanging with the Medical physics group of the National University proposed a model for DNA molecules in Geant4-DNA \cite{chauvie2007geant4}. This was a remarkable advancement for the software. In the proposed model, the DNA is parameterised under geometrical conditions. Fig. \ref{2} shows an example of the proposed model for DNA molecules. This model presents a three-level organisation for the DNA which includes the double helix, the nucleosome and the chromatin fibers. Nowadays there are many versions of DNA geometric models implemented in Geant4-DNA, which are freely available.

\begin{figure}
\centering
\includegraphics[width=0.4\textwidth, height=0.2\textheight]{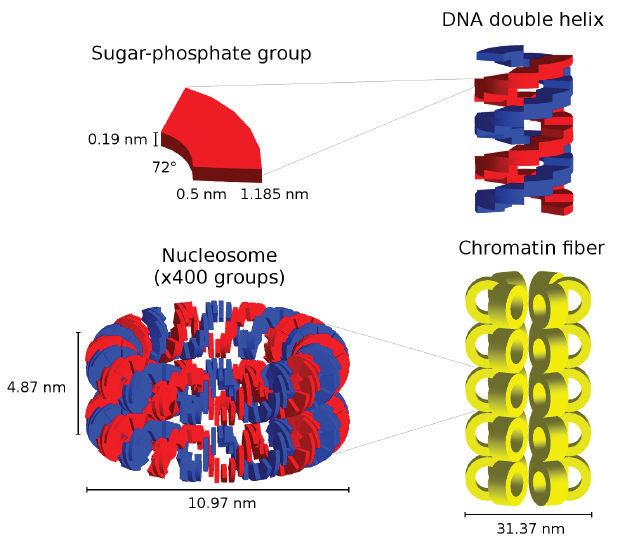}
\caption{Model of three organization levels of the DNA developed in Geant4 software \cite{chauvie2007geant4}.}
\label{2}
\end{figure}

\subsection{MATLAB Simulation}

In this paper, it is presented the simulation for the interactions between X-rays and DNA molecules considering the structure and atomic composition of DNA, the advantage of this software is the control over the geometrical and spatial components of each atom and each bond that compose the molecule of interest. It is also possible to extend the atomic composition based on corrupted cases where the molecule is embedded in different chemicals. The spatial control of specific atoms is a component that must be taken into account in the quantification of the damage where I do not only want the distribution of atoms but the deterministic positions. The DNA molecules are created according to spatial coordinates of the atoms that are joined through vectors which emulate the bonds. The spatial distribution of the electrons generated after the X-rays radiation of $17.6 \, \text{KeV}$ in water is obtained based on Geant4. Randomness is assumed for the X-rays with different origins in the three dimensional space which are generated through Monte Carlo simulation. The spatial distributions of electrons that is generated are spatially included in Matlab based on the saved results of Geant4.
It is taken into account a relevant factor to detect damage which is sequence of the DNA molecules since in the simulations, the variation of the sequence and size of DNA molecules is detectable. Hence, this advantage opens the door for researchers to study the effect of bases contents on the level of vulnerability of DNA. This characteristic makes possible to detect base damage and not only DNA strand damage. The other track Structure codes also consider DNA size and sequence but are not atomic specific in space, preventing the considerations on spatial structure transformation.\\ \\
The results presented in this paper consider $10,000$ pair bases with a total of $634,983$ atoms. The number of bases could be increase in the simulation however, this is the maximum considered number due to the computational power constraint. This software is not implementing plasmids or full cell nuclear DNA as it could be found in other softwares but it creates from scratch the spatial components of the double helix. Computations are easily implemented thanks to the vectorisation of the atoms and its bonds, allowing linear algebra. This is a more deterministic approach in comparison to a probabilistic framework used by the others. 

\subsection{Experimental Verification}

In order to observe the damage in the DNA after its respective irradiation, I use the Atomic Force Microscopy (AFM) which works based on hitting an atomic tip with the DNA samples and maintaining a constant force between the tip's atom and the DNA sample's atom. It sweeps fast the samples with an oscillation of $68 \, \text{kHz}$ \cite{eaton_atomic_2010}. This frequency is provided by a piezoelectric material with high precision. Changing the location of the tip relative to sample's atoms, provides three spatial data points to create a surface which approximates the appearance of the sample in the atomic scale. The process of microscopic visualisation is not simple however, usually the first obtained image form AFM can be considered as the image of a strand of DNA. The preparation protocol of the sample and its analysis is well documented and standardised \cite{cerreta_fine_2013}. The DNA sample should be reduced to the adequate concentrations. The sample is placed over a negatively charged surface called MICA. Since the DNA is also negatively charged, a positive buffer is added which serve as interface between both phases. The sample could be dry or in water \cite{eaton_atomic_2010}.

\section{Methodology}
\subsection{Simulation}
\subsubsection{DNA Creation}

\begin{figure}
\centering
\includegraphics[width=0.3\textwidth, height=.2 \textheight]{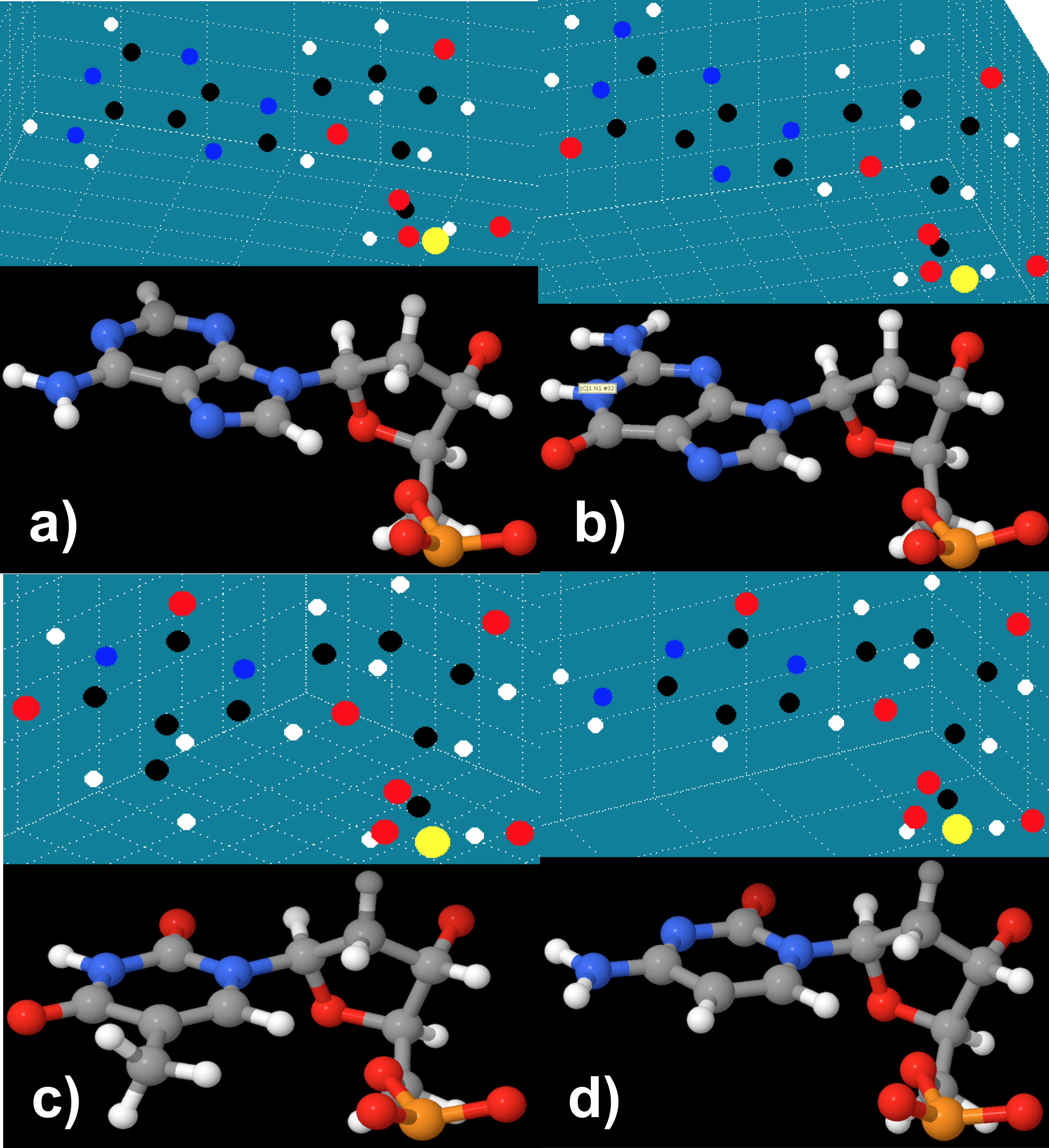}
\caption{Molecules visualisation in Matlab: a) Adenine, b) Guanine, c)
Thiamin y, d) Cytosine. The obtained molecules in the PDB are in the background.}
\label{f:bases}
\end{figure}

The key part of the simulations is creating a double DNA helix. In order to create the DNA molecules in a realistic way, I need information about the location of DNA atoms and its spatial coordinates. This information can be obtained from the \emph{Protein Data Base} (PDB). It is a file which provided the spatial coordinates of $637$ atoms which compose the double helix with $10$ base pairs and it is beneficial for the DNA visualisation. The PDB file can only be used for visualisation and data extraction is difficult. This is the main disadvantages of PDB files. For solving this difficulty, the extraction of the spatial data from this file is required and then use it for the simulation in MATLAB. The data extraction from the PDB file is done with the aim of introducing all types of sequences and achieving the correspondence helix. Hence, I can compare different DNA sequences in terms of their sensitivity to the damage. The PDB file is organised such that each atom has an atomic number. Then, in order to identify different bases, I assign a number to each base in alphabetic order. Next, this file is exported to MATLAB. Then the atoms which are a part of each molecule are identified with the aim of organising, comparing, and finding their pattern in the molecule structure.\\ \\
In order to identify the common bonds in the structure. Three more important bonds are use for comparison between molecules according to their order in the strand. This is done to obtain the molecules' rotations when they shape the helix. Therefore, three vectors are required to store the information regarding the line between bonds. These vectors are compared between the molecules of each helix through the following rotation matrix:
\begin{equation}
R=
\begin{bmatrix}
C_\psi C_\theta& C_\psi S_\theta S_\phi-S_\psi C_\phi& C_\psi S_\theta C_\phi+ S_\psi S_\phi\\
S_\psi C_\theta& S_\psi S_\theta S_\phi+C_\psi C_\phi& S_\psi S_\theta C_\phi- C_\psi S_\phi\\
-S_\theta & C_\theta S_\phi & C_\theta C_\phi
\end{bmatrix},
\end{equation}
where $\phi$, $\theta$ and $\psi$ are the rotation angles about the axis X, Y, Z, respectively. The $C_\phi$, $C_\theta$, $C_\psi$ are cosines of these angles and $S_\phi$, $S_\theta$, $S_\psi$ are their sines. The rotation matrix and position vectors determine the location of molecules in the correct place and in the correct direction. This is done for each nitrogen base with the associated location.
Fig. \ref{f:bases} presents a visualisation obtained from the PDB data file. 
The four molecules which constitute the DNA were created through the PDB file in order to develop the simulations where the desired sequence is the input and it can be created according to the atoms spatial locations. The results of the modelled molecules can be seen in Fig. \ref{f:bases}.

\subsubsection{Interactions of X-Rays in Water}

In order to simulate the effect of X-rays on the DNA, I use a Monte Carlo simulation for the rays trajectories since the trajectories of radiations are not predictable. Fig. \ref{f:DNAxb} shows random trajectories for X-rays where the size of the strand is double and these trajectories are corresponding to the photons that hit the DNA and as a product of this hitting, electrons are generated in the proximity of DNA.
The X-rays have a remarkable property. They can generate a large amount of electrons when they traverse through a water-based medium since their energy is higher than the ionising energy of water. This fact has as consequence that there is a larger amount of cascade reactions due to the X-rays radiation which is manifested in the easy production of free radicals and its effect such be included in the simulation given that they are responsible for most of the DNA damage.
The spatial distribution of electrons in water is obtained with the software Geant4. This software is useful to provide experimental and theoretical information of the spatial distribution of the electrons which are generated in the proximities of the X-rays trajectories. The theoretical backgrounds for simulating this process have been developed in many papers and they are not the scope of this research, more detailed information can be found in \cite{Lampe2018}.
After adding information about the spatial distribution of electrons into MATLAB, it is possible to simulate damage. Moreover, the simulation includes the effect the large amount of particles which are generated around the X-ray bean.

\subsubsection{Visualisation and DNA Damage Quantification}

In order to quantify the damage that the X-rays generate, it is necessary to save the state of the DNA and the distributions of its atoms after each iteration. Hence, it is possible to keep track of the reactions in the DNA structure and the events after the collisions with X-rays or with electrons. This information is then used to analyse the SSB and DSB damage after exposition to different radiation intensities. There is a threshold dependent on the number of interactions and not in energy, given that energetic considerations will require potential details of atomic forces that are not under the scope of this paper, for more detail regarding energetic thresholds for damage quantification, please refer to \cite{Alizadeh2011}. \\ \\ 
For the purpose of this paper, the damage is quantified based on a threshold in the counting of allowed interactions and when the distance between the electron and each DNA atom is less than two Van Der Waals radius, a count in the number of damage events increases. Then, based on the analysis of the location of the hits might or might not result in the disappearance of crucial bonds of the helix, given that are sensible to base damage. The damage is quantified when a DSB is detected instead of a SSB. A DSB damage occurs when the nucleotides faced right in front of each strand have a damage and this condition is checked every iteration. Two SSB damage is accounted as a DSB damage when they have the same locations in opposite strands of the helix.

\subsection{Experimental Procedure}
\subsubsection{DNA Extraction}

It has been employed a protocol for blood DNA extraction which uses a blood sample of a health volunteer. After the DNA extraction I divided the sample into seven aliquots. Each one has a concentration equal to $20\, \frac{ng}{\mu L}$ with a volume equal to $50\,\mu L$. They were immersed in water to simulate the cell conditions since it is mainly composed of water. The exploited water was distillate, deionized and pure. These aliquots are irradiated with the following radiation doses: $0.006 \, \text{Gy}$, $0.025 \,\text{Gy}$, $0.08 \,\text{Gy}$, $1 \,\text{Gy}$, $5 \,\text{Gy}$ and $30 \,\text{Gy}$, where Gray is absorption of 1 Joule of ionising radiation per kilogram of mass and this is the most used unit in radio-biology. One of the samples which is not irradiated, while it is under the same conditions of transportation, movement and temperature as the irradiated samples, is denoted as \emph{negative control} sample. The negative control sample is the bases for the comparisons to show how much damage was generated in the DNA. It is worth mentioning that the aliquots are kept in ice during the irradiation process and then they are conserved to $4 ^{\circ}C$.

\subsubsection{X Rays-DNA Interaction}

\begin{figure}
\centering
\includegraphics[width=0.4\textwidth, height=.2 \textheight]{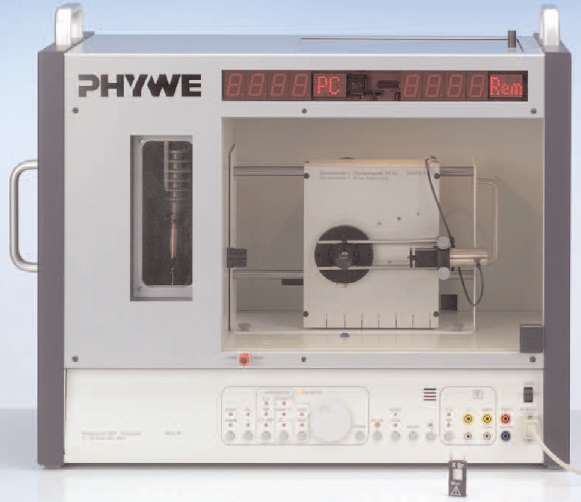}
\caption{X-rays module employed for irradiation.}
\label{f:phi}
\end{figure}

\begin{figure}
\centering
\includegraphics[width=0.4\textwidth, height=.2 \textheight]{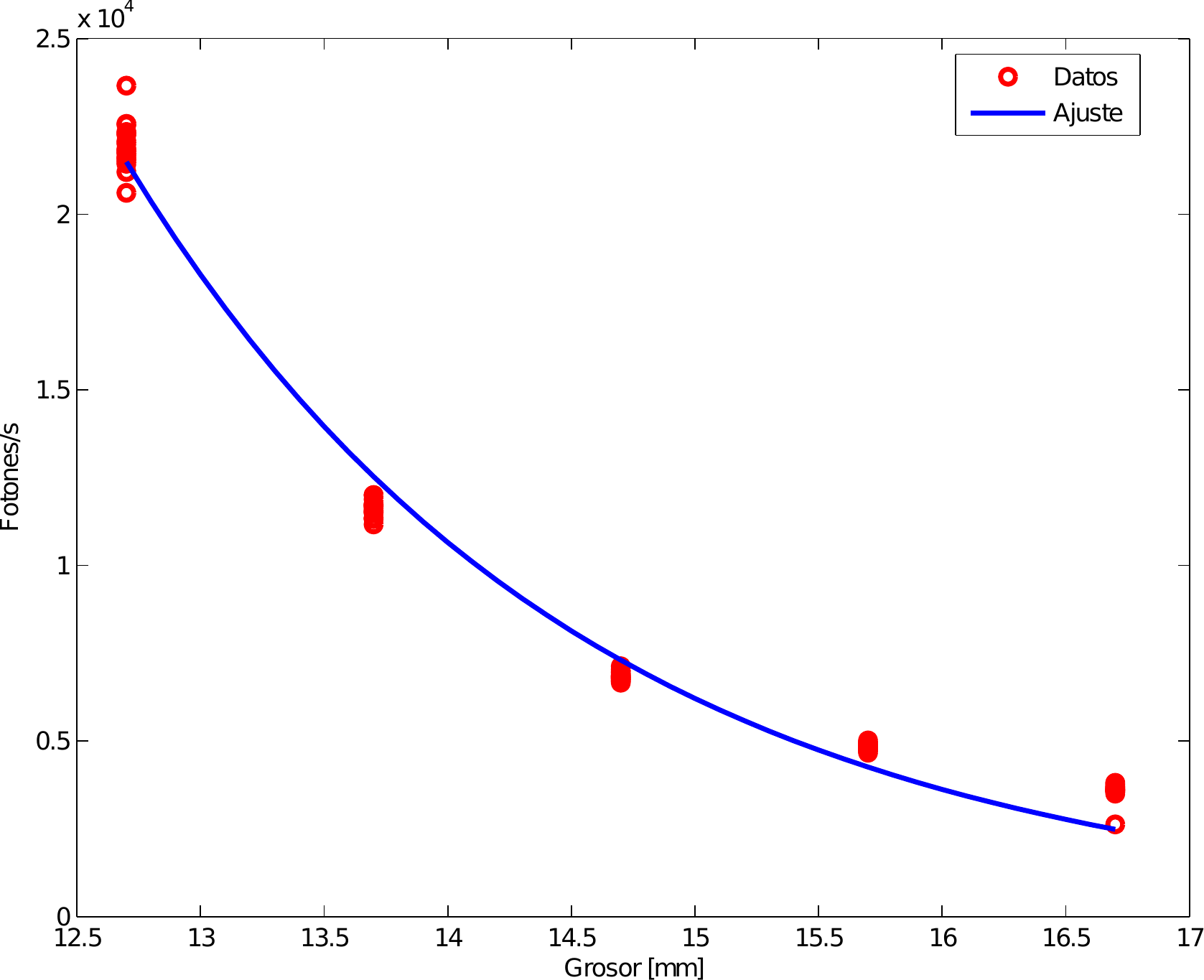}
\caption{The relation between the number of photons and the thickness of Aluminium plates and an exponential regression.} 
\label{f:exp}
\end{figure}

Each one of the six aliquots were irradiated by a basic unit of X-rays with $35 \,\text{KeV}$ in different Grays (Gy). The maximum peak of the work function of this unit is $17.6 \,\text{KeV}$. This is the real energy that the samples are received. X-rays were generated by a PHYWE machine as it is shown in Fig. \ref{f:phi}.
The selected radiation values are obtained as follows: $0.006 \,\text{Gy}$ is obtained with an exposition time equal to 2 seconds, $0.025 \,\text{Gy}$ with 8 seconds, $0.08 \,\text{Gy}$ with 27 seconds, $1 \, \text{Gy}$ with 5 minutes 35 seconds, $5 \, \text{Gy}$ with 27 minutes 58 seconds and $30 \, \text{Gy}$ with 2 hours 47 minutes 50 seconds. The exposition times are derived through a calibration method. This method shoots a beam towards aluminium plates with different thickness and takes the photons and counts by a Geiger tube. The relation between the number of photons and the thickness of aluminium plates is studied. Then, an exponential regression is used in order to obtain the total number of photons corresponding to an exposition without an aluminium plate. The results of the exposition process are presented in Fig. \ref{f:exp}.
The power is calculated as the number of photons per second. The mentioned power is the absorbed energy per second. This is obtained by the Molybdenum work function since the X-rays emission tube is made of Molybdenum and its electrons have a maximum of energy equal to $20\, \text{KeV}$. As mentioned before, the Gray unit needs the mass of the exposed sample. Hence, it is necessary to find the DNA mass. Thus, to derive the mass of DNA, I use the volume and the density of the sample. Therefore, obtaining the exposition time for all desired irradiation values as follows:
\begin{equation}
T_{\text{exposition}}=\text{Gy}\cdot \frac{m_{\text{DNA}}}{\text{Potential}},
\end{equation}
where $\text{Gy}$ is the desired irradiation value and $m_\text{DNA}$ is mass of DNA. I selected these exposition values since they are the most used values in radiotherapy applications. Moreover, it is desirable to have a logarithmic sampling to generate converging results. The amount of $30 \,\text{Gy}$ is applied to patients in radiotherapy; however, this is an accumulative radiation which is obtained after many sessions. The aim of this procedure is the healthy cells which are affected by an irradiation and have enough time to repair. On the other hand, since carcinogenic cells are repaired at a slower rate, the time between irradiations is not enough for them to repair, and thus, they are eliminated. 

\subsubsection{Damage Visualisation and Quantification. Atomic Force Microscopy (AFM)}
It was used a DNA sample preparation protocol which is a special procedure for the AFM. It requires a buffer preparation with $NiCl_{2}$ which is useful to fix the DNA. This protocol uses a surface with minimum roughness in atomic scale which is denoted as MICA. It has a negative charge similar to the DNA. Thus, the $NiCl_{2}$ is used to create an interface between the MICA and the DNA since it has positive charge \cite{eaton_atomic_2010}. Another relevant aspect for the visualisation in the AFM is the DNA concentration in the sample preparation which should be $5\,\frac{\mu \text{g}} {\text{mL}}$ and then $1\,\frac{\mu \text{g}}{\text{mL}}$ . Finally, $37 \,\mu \text{L}$ are added to the mixture of DNA and buffer over the MICA which acquires the shape of meniscus.\\ \\
This protocol has the option of making the sample in dry or liquid conditions. In this research, the sample was prepared in dry condition. It is worth mentioning that the preparation of the sample is the most important factor for the success of the microscopy and isolation condition is desirable for the preparation process. For the damage quantification, the irradiated samples are compared with the negative control sample in terms of their fragment size.
\section{Results}
\subsection{Simulation Results}
\begin{figure}
\centering
\includegraphics[width=0.4\textwidth, height=.3 \textheight]{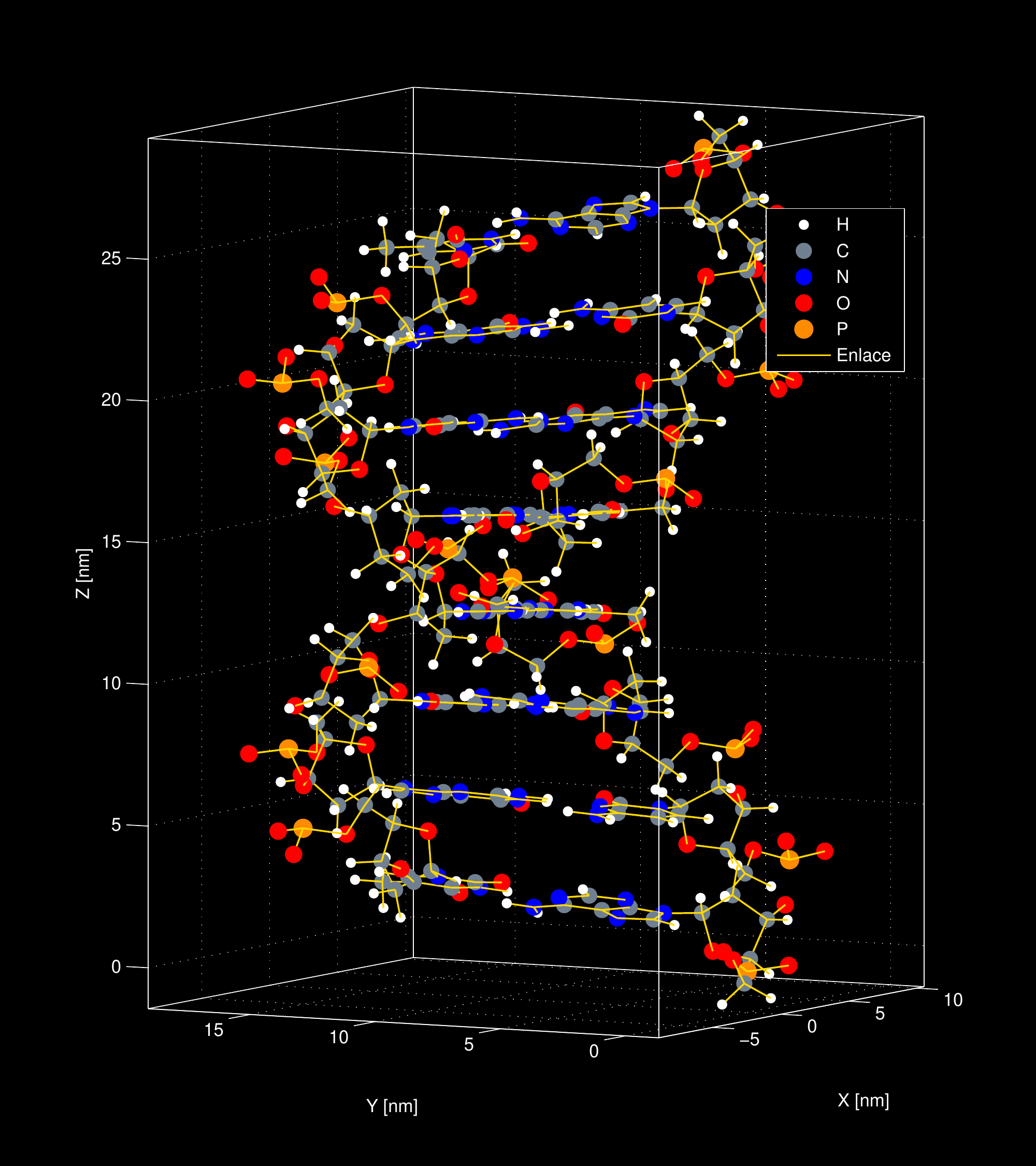}
\caption{A short DNA sequence obtained by the developed software.} 
\label{f:DNAxa}
\end{figure}

As the first result, it can be seen that the proposed software can generate single-helix and double-helix DNA with any length and any desirable sequence. This software creates the sequence with the respective atoms associated to the nucleotides and their respective bonds as shown in Fig. \ref{f:DNAxa}.

\begin{figure}
\centering
\includegraphics[width=0.4\textwidth, height=.3 \textheight]{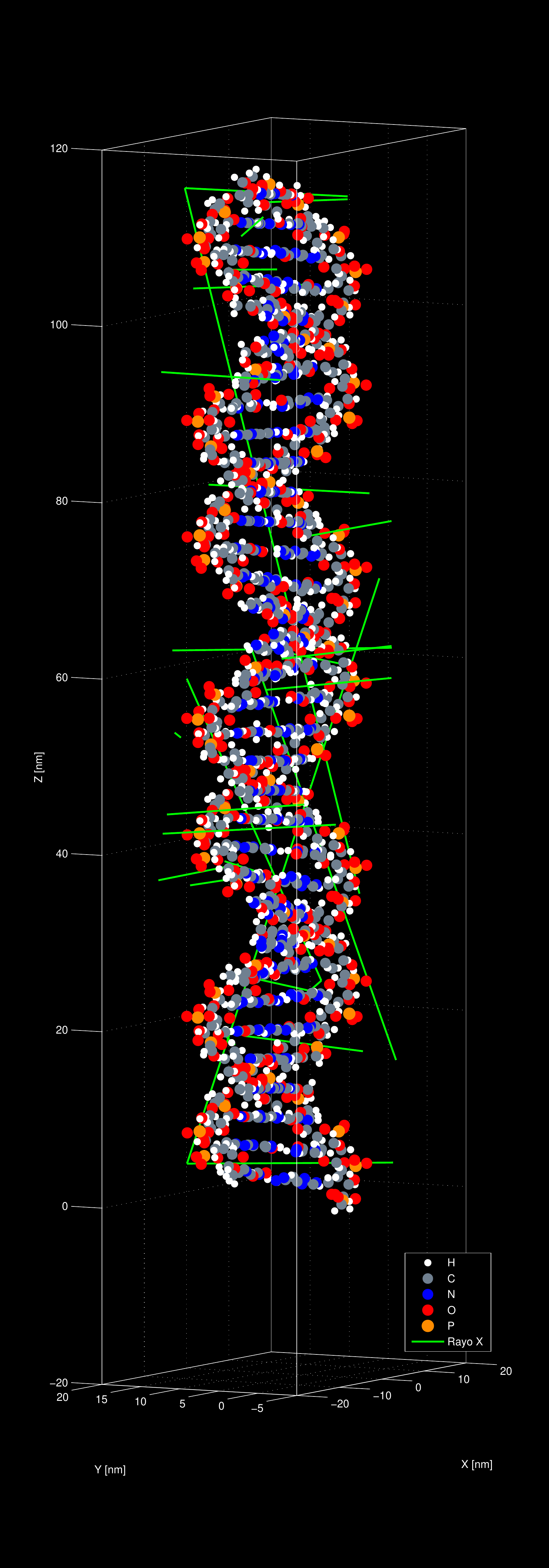}
\caption{Random trajectories generated by a Monte-Carlo simulation.} 
\label{f:DNAxb}
\end{figure} 

\begin{figure}
\centering
\includegraphics[width=0.4\textwidth, height=.3 \textheight]{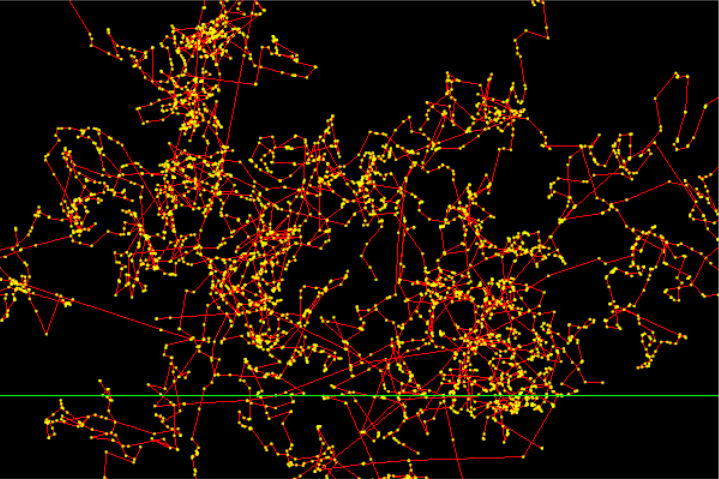}
\caption{ Free radicals generation simulation in Geant4, for a photon energy of $20\, \text{keV}$.}
\label{f:rad}
\end{figure}

\begin{figure}
\centering
\includegraphics[width=0.4\textwidth, height=.3 \textheight]{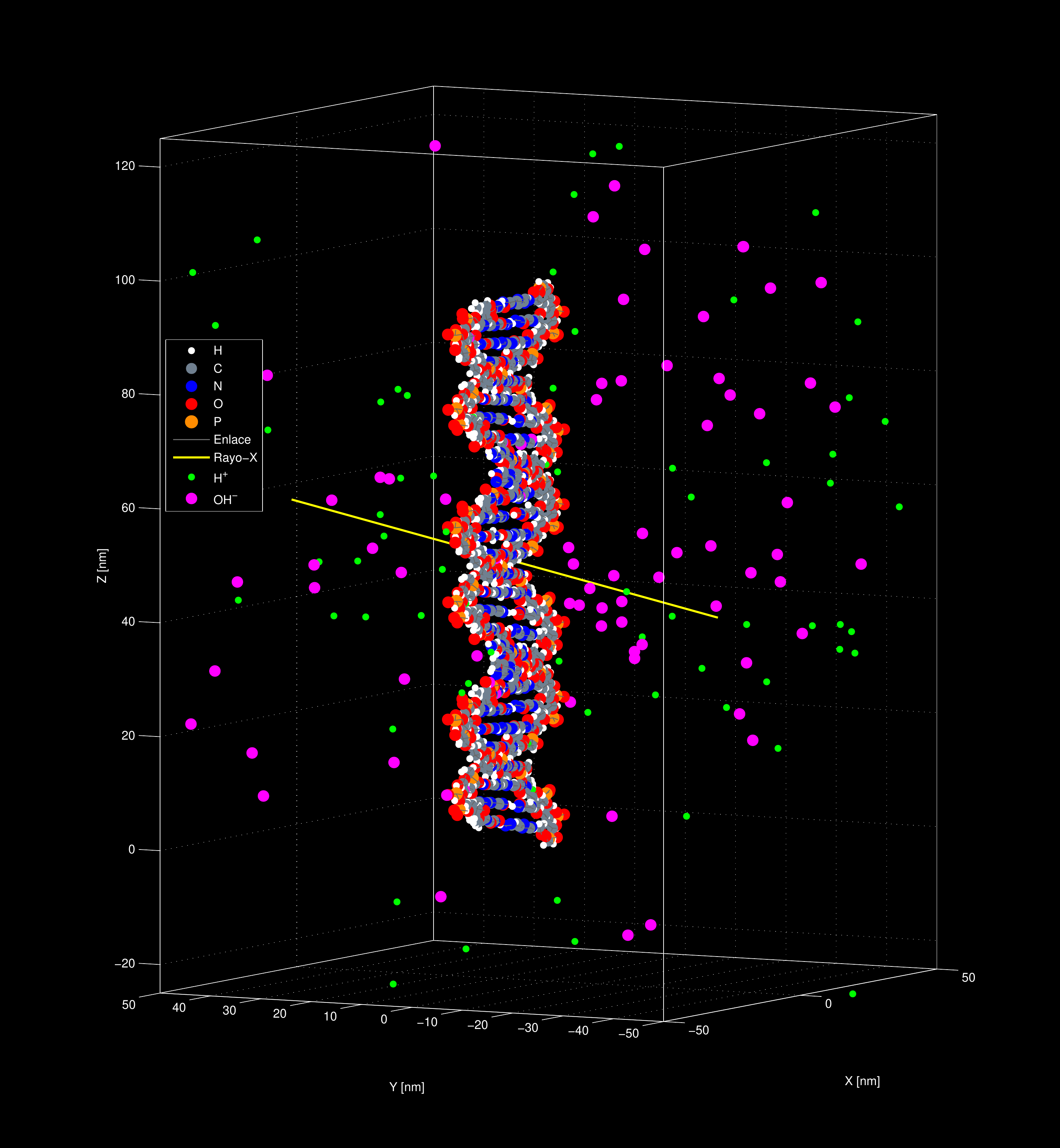}
\caption{The spatial distribution of particles that are generated in the proximities of the trajectories of the X-rays.}
\label{f:DNAelectrones}
\end{figure}

\begin{figure}
\centering
\includegraphics[width=0.4\textwidth, height=.3 \textheight]{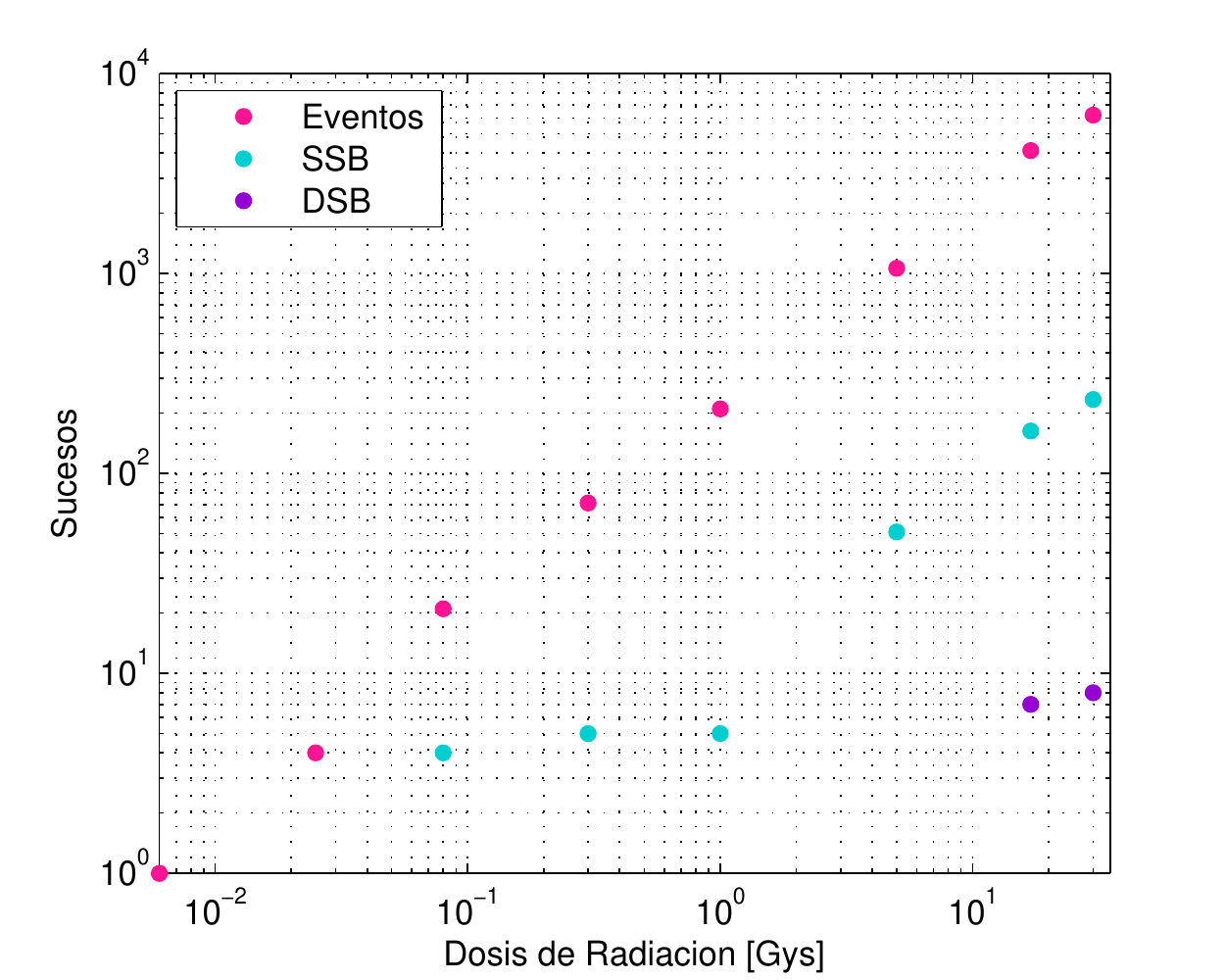}
\caption{Te number of events (SSB and DSB) for different values of radiation (Gy).}
\label{f:cuantificacion}
\end{figure}

Many rotation matrices are obtained which are corresponding to the rotations that happen between different molecules when I move along the DNA strand. According to the obtained matrices, I can conclude that in the DNA each molecule rotate 36$^\circ$ counterclockwise when I move along the helix, and thus, I have $\phi=0$, $\theta=0$ and $\psi=36^\circ$. The rotation angles are similar for all molecules along the DNA. Thus, when I observe the DNA, the bases are parallel. This is an indication that the basis only can rotate over one plane. This result agrees with what is reported in \cite{sinden_dna_1994}.\\ \\
This software can create any type of DNA sequence. An example of DNA sequence is shown in Fig. \ref{f:DNAxa} which is corresponding to the sequence \texttt{ACGTACG}. The sequences can be freely chosen in terms of length and content. This is a valuable characteristic of this software, in comparison to PARTRAC which also have geometrical considerations of the interaction but is not considering the atomic detail resolution, on the other hand, Geant4-DNA has the atomic resolution but its not encoding the structure geometrically, which is important in the quantification of the damage given that after each iteration will create new conformations that can expose or protect the strand from the cascade of interactions. This computation is possible by linear algebra with the rotation matrix and the position vectors. The software described in this paper, takes advantage of this geometrization and compares the vulnerability based on the level of nucleotide content.\\ \\ 
For example, it is possible to conclude that the sequence with a high Guanine-Cytosine content is stronger against damage than a sequence with a low Guanine-Cytosine content. Fig. \ref{f:DNAxb} shows a visualisation where DNA is damaged in a direct way by the generated random radiations. The trajectories of photons are indicated by the straight lines. It is worth mentioning that all of trajectories do not create a SSB or DSB because it is imposed a constrain in atomic distance in order to count as an interaction.\\ \\
Implementation of all reactions in MATLAB takes a long time and it is out of the reach of this research. Therefore, it is possible to use the results of Geant4 to obtained the effect of X-rays on DNA embedded in water and the output of the distribution of free radicals around one X-ray trajectory and not the point to point position of the free radicals generation. This is due to the fact that there is a large amount of particles created after X-rays traverse a medium of water and it would require a lot of memory to store the specific positioning of the free radicals. Fig. \ref{f:rad} shows the complexity of this process. The resulted products of the X-ray trajectory are marked by thin lines and it can be observed an enormous amount of free radicals (tiny particles) which are generated by the X-ray (the straight and thick line). The results are saved and used for different X-rays simulations. Thus, by employing superposition property, it is possible to generate the visualisation of the generated particles for different X-rays initial conditions. Fig. \ref{f:DNAelectrones} shows the DNA helix, the X-ray and created particles together with their interactions in one of the simulation scenarios.
Finally, the number of interactions is counted and number of SSB and DSB damage is quantified. The simulation scenarios are implemented for different radiation values and for an helix with $10000$ pairs bases. The simulation scenario requires a lot of computational power. It takes about eight hours per run and the DNA is constantly changing according to its interaction effects. The trajectories of X-rays are created by Monte-Carlo simulation containing a large number of photons of high energy. The total number of events are shown in Fig. \ref{f:cuantificacion} for different radiation doses. This figure indicates a linear behavior for number of events by increasing radiation. As it was expected the DNA -electrons and DNA-X-ray reactions are increased by increasing radiation. In can be observed that there is no SSB for $0.006 \,\text{Gy}$ and $0.025 \,\text{Gy}$ doses since they are weak. Moreover, there are about 5 damage for $0.08 \,\text{Gy}$, $0.3 \, \text{Gy}$, and $1 \, \text{Gy}$ doses.
The SSB damage increase exponentially for $1 \,\text{Gy}$, $5 \,\text{Gy}$ and $30 \,\text{Gy}$. For $30 \,\text{Gy}$, the number of events is $6000$ and there are many interactions in this situation. Hence, the DNA becomes weaker and the number of damage grows. Moreover, the DSB appear only for $17 \,\text{Gy}$ and $30 \,\text{Gy}$. The number of DSB damage is $8$ for $30 \,\text{Gy}$.
The results in Fig. \ref{f:cuantificacion} confirm the reported results in the literature. When the number of event is in order of $10^{9}$, the number of DSB is in order of reported results in \cite{friedland_track_2011}. Thus, it shows the accuracy of the simulations. It is important to notice that, as expected, the number of SSB and DSB is lower compared to other simulations, because of the dependency of the damage with the energy of the X-ray. This simulation is done with energies on the order of KeV, this energy was set given that the experiment was done for this same energy but this parameter can be easily change in order to make comparisons with other simulations which usually use more realistic therapeutic energies on the MeV range. \\ \\
Difference in the MeV regime produce damage behavior changes from linear to exponential and poisson. This effect can also be found in \cite{Campa2005}. Furthermore, the number of photons which interact with the DNA is determined according to the experimental results. However, this number can be modified when the number of interacting photons is increased and consequently the damage probability also increases. The number of interacting photons is constrained to the computational power.

\subsection{Experimental Results}

\begin{figure}
\centering
\includegraphics[width=0.4\textwidth, height=.3\textheight]{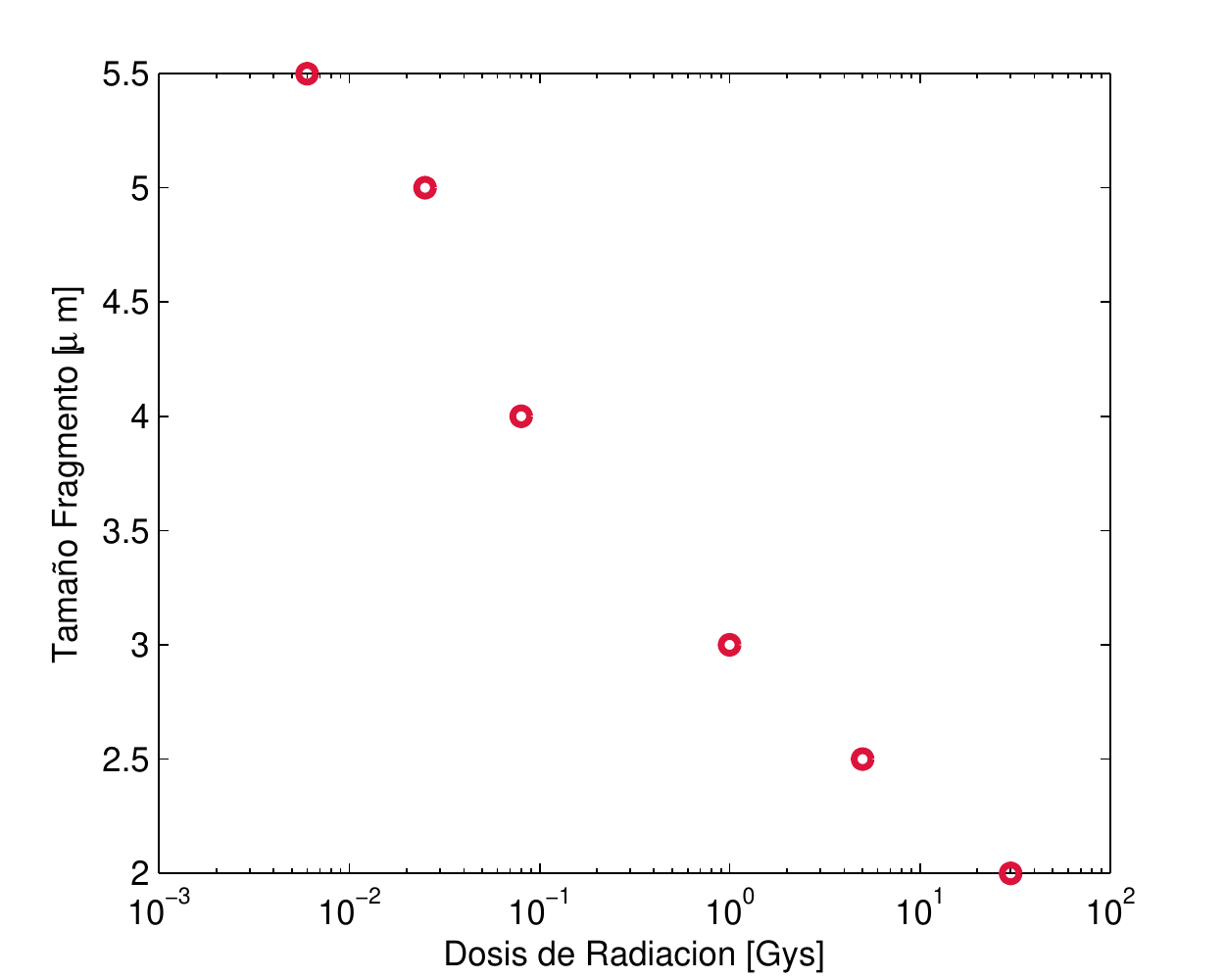}
\caption{The DNA fragment size for different radiation doses.}
\label{f:AFM}
\end{figure}

\begin{figure}
\centering
\includegraphics[width=0.4\textwidth, height=.3\textheight]{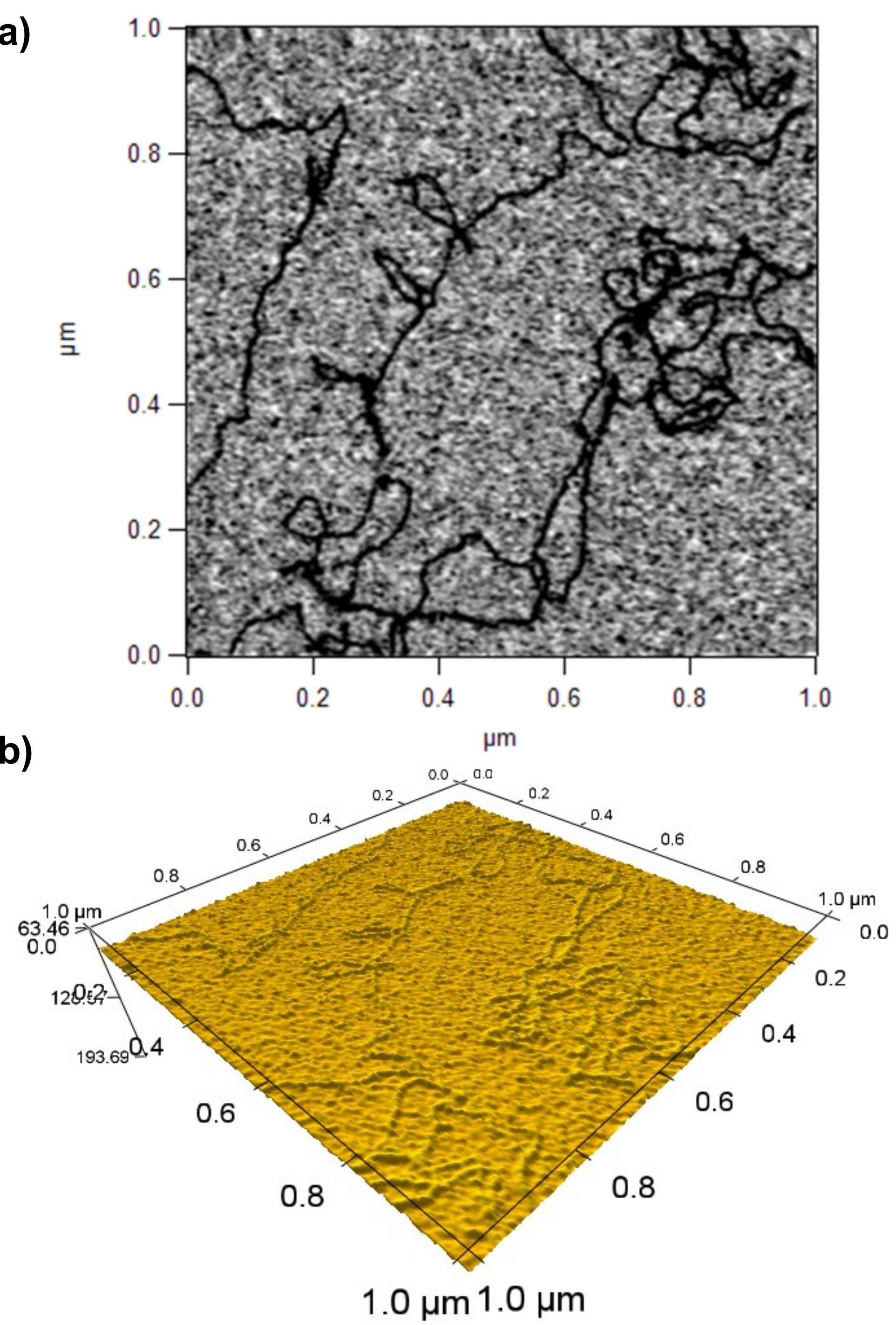}
\caption{Negative control results: a) 2D view of the AFM, b) 3D view of the AFM.}
\label{f:control}
\end{figure}

\begin{figure}
\centering
\includegraphics[width=0.4\textwidth, height=.3\textheight]{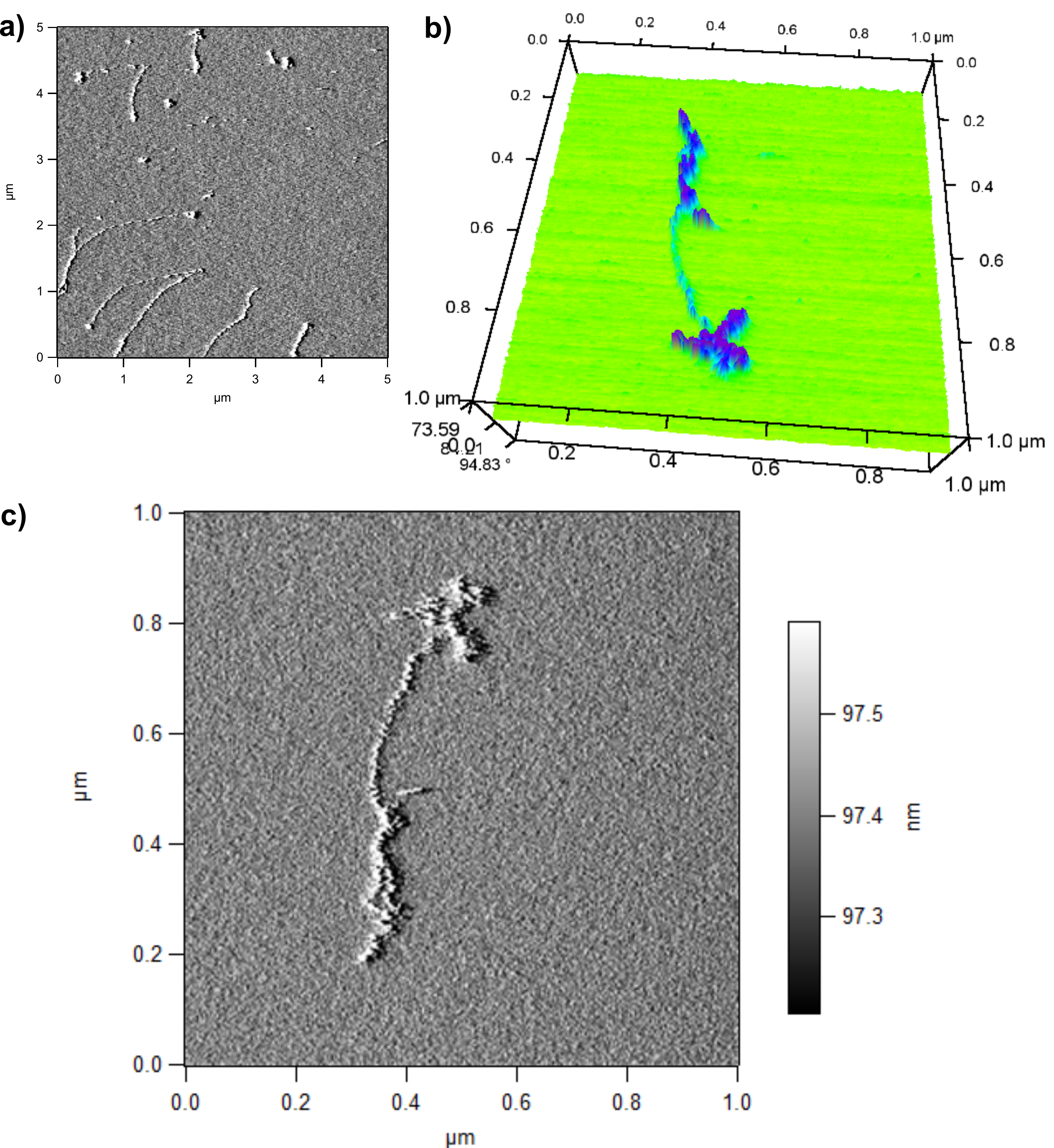}
\caption{Irradiated sample with $0.006 \,\text{Gy}$, a) 2D view of the AFM to 5$\,\mu m$, b) 3D view of the AFM to 1$\, \mu m$, c) 2D view of the AFM to 1$\,\mu m$.} 
\label{f:0006gy}
\end{figure}

\begin{figure}
\centering
\includegraphics[width=0.4\textwidth, height=.3\textheight]{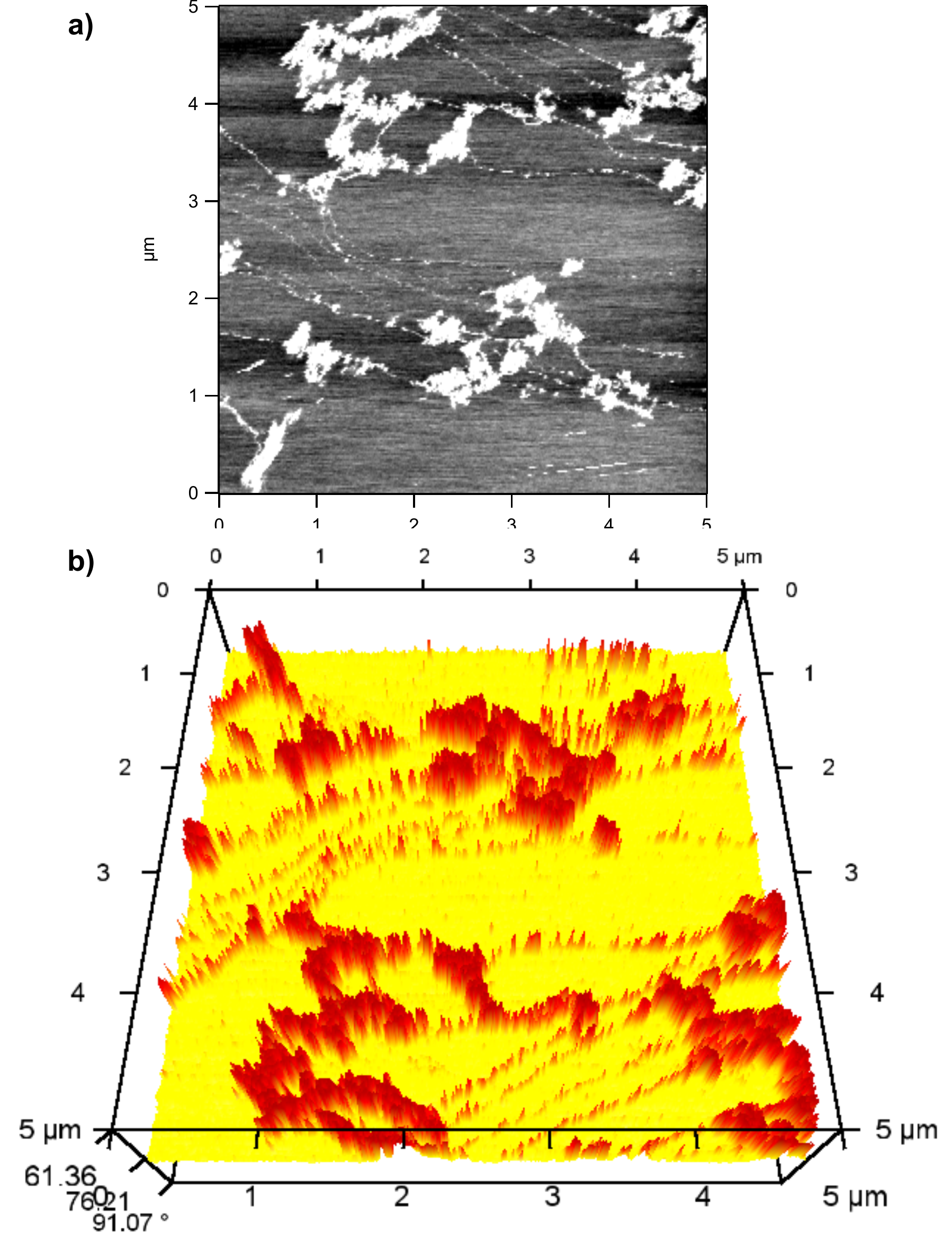}
\caption{ Irradiated sample with $0.025\, \text{Gy}$, a) 2D view of the AFM to 5$\,\mu m$, b) 3D view of the AFM to 5$\, \mu m$.}
\label{f:0025gy}
\end{figure}

\begin{figure}
\centering
\includegraphics[width=0.4\textwidth, height=.3\textheight]{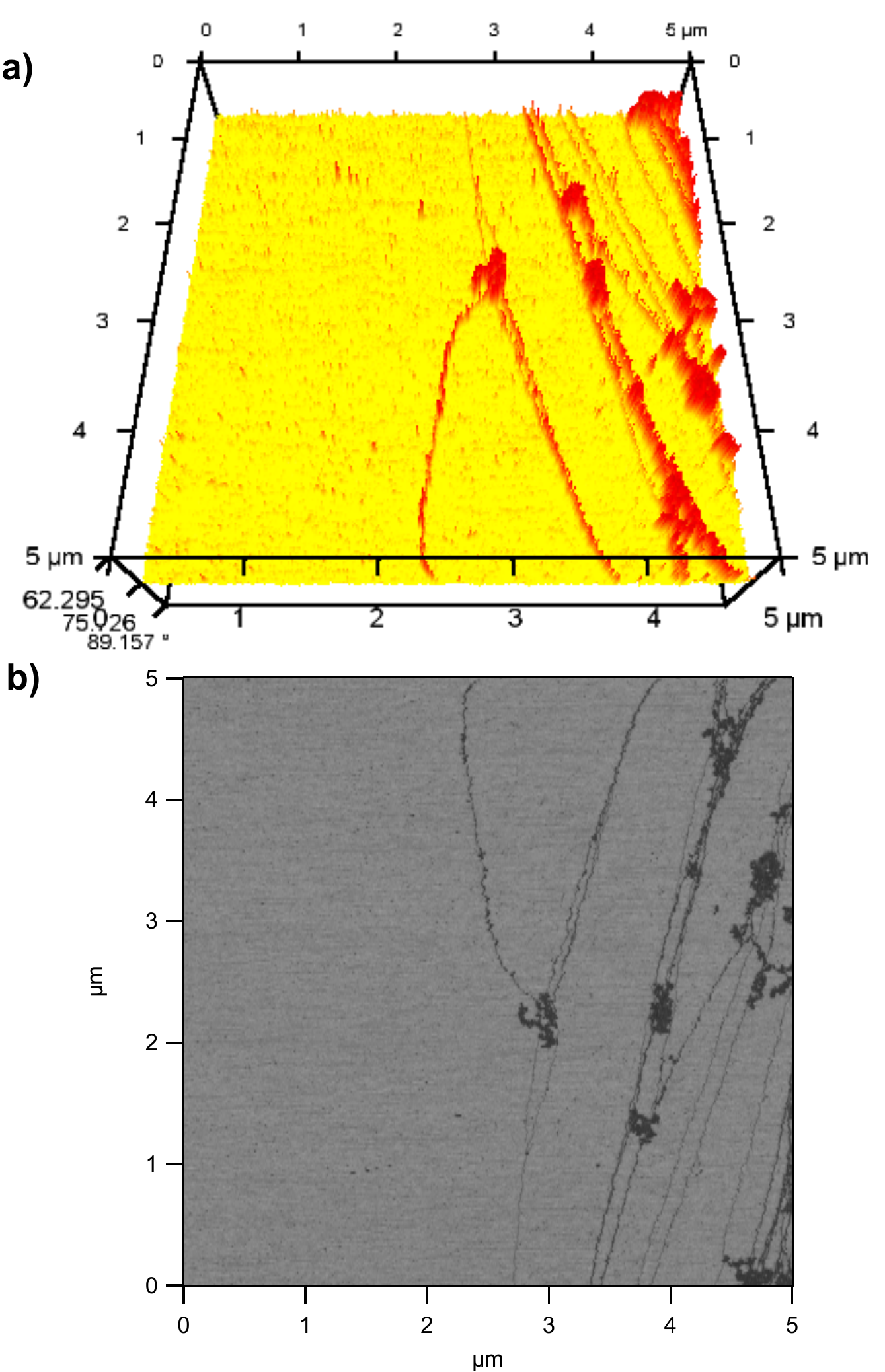}
\caption{Irradiated sample with $0.08\, \text{Gy}$, a) 3D view of the AFM to 5$\,\mu m$, b) 2D view of AFM to 5$\,\mu m$.}
\label{f:008gy1}
\end{figure}

\begin{figure}
\centering
\includegraphics[width=0.4\textwidth, height=.3\textheight]{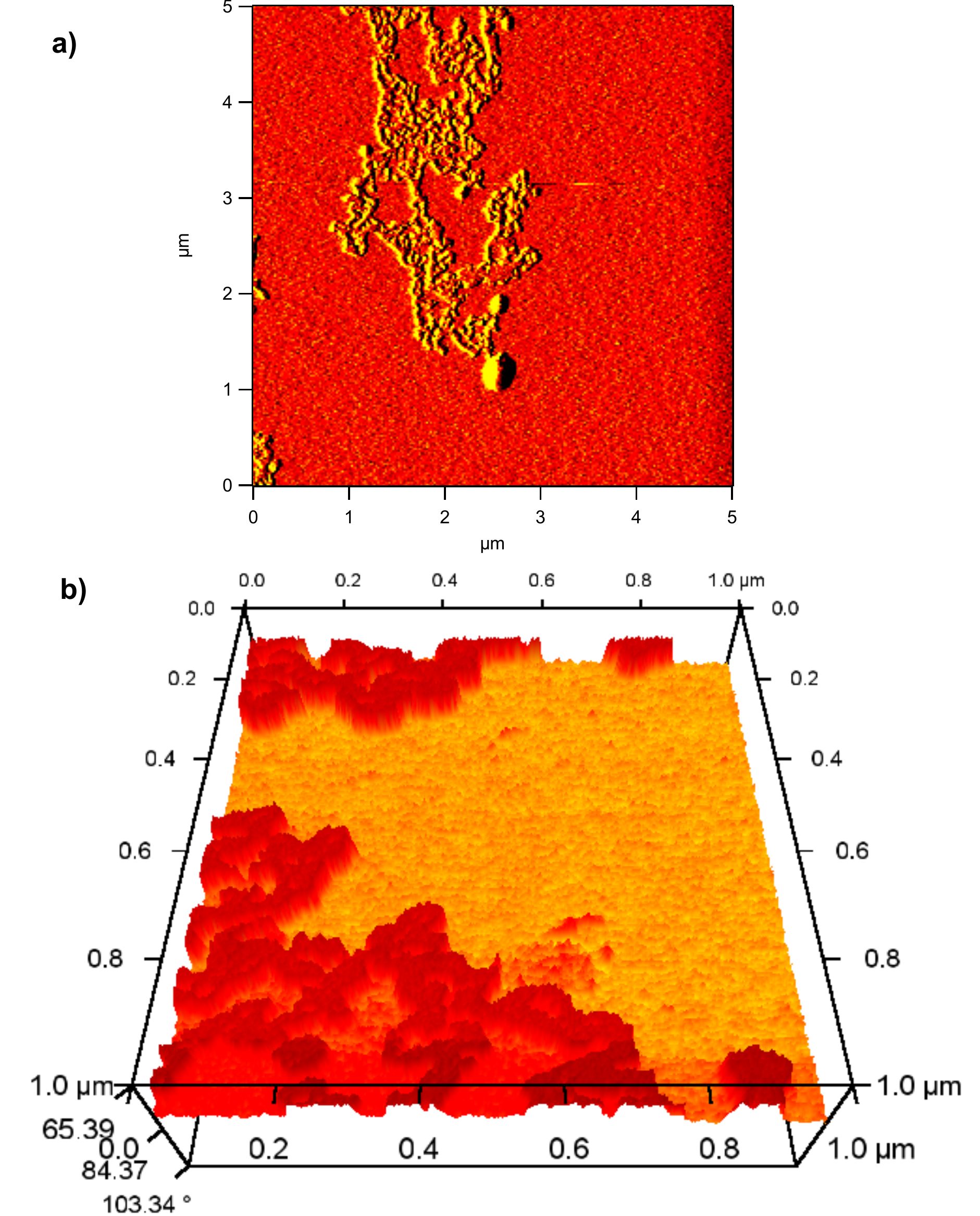}
\caption{ Irradiated view with $1\, \text{Gy}$, a) 2D view of the AFM to 5$\, \mu m$, b) 3D view of the AFM to 1$\,\mu m$.} 
\label{f:1gy1}
\end{figure}

\begin{figure}
\centering
\includegraphics[width=0.4\textwidth, height=.3\textheight]{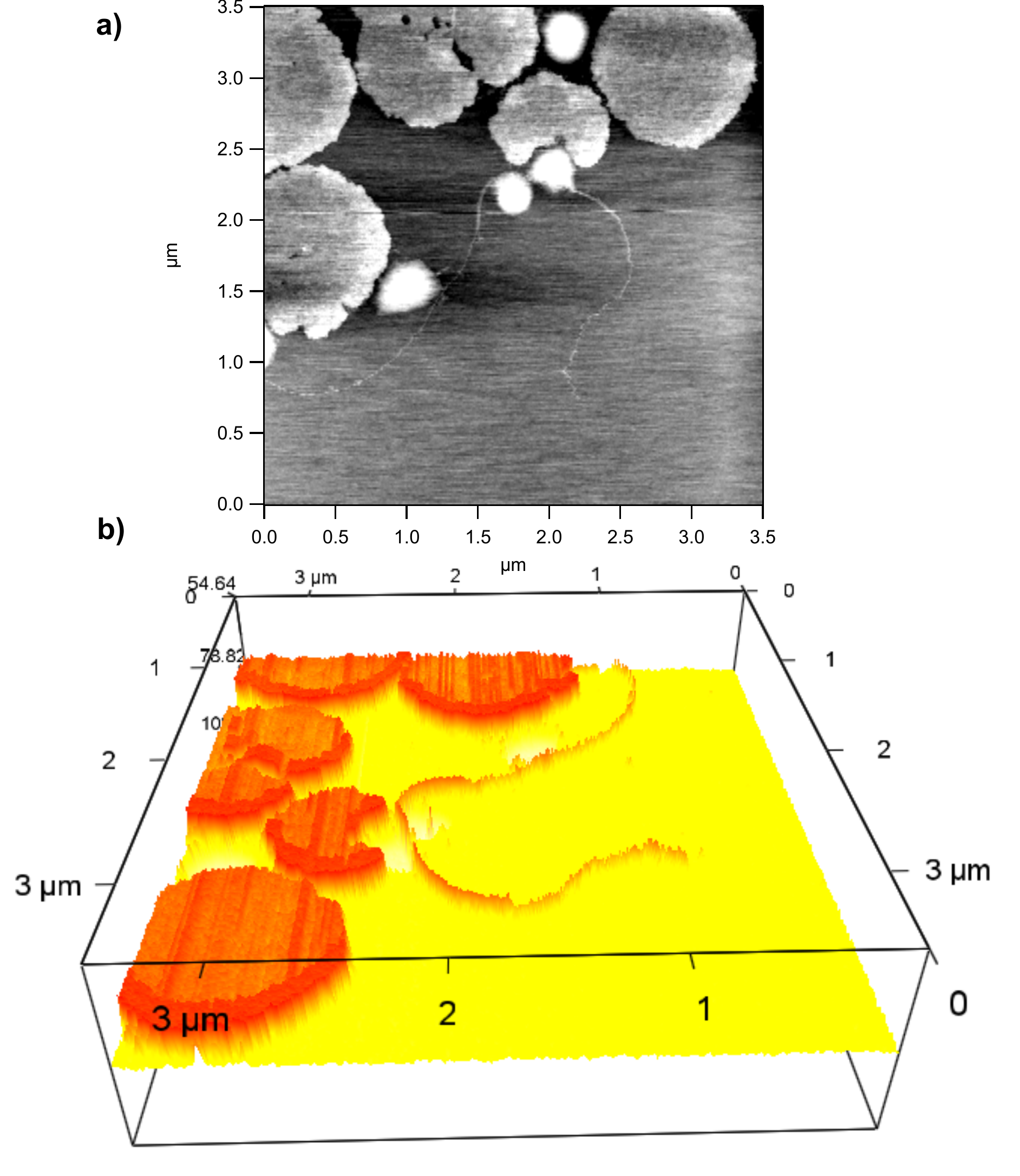}
\caption{Irradiated sample with $5 \,\text{Gy}$, a) 2D view of the AFM to $3.5$$\, \mu m$, b) 3D view of the AFM to 3$\, \mu m$.}
\label{f:5gy1}
\end{figure}

\begin{figure}
\centering
\includegraphics[width=0.4\textwidth, height=.3\textheight]{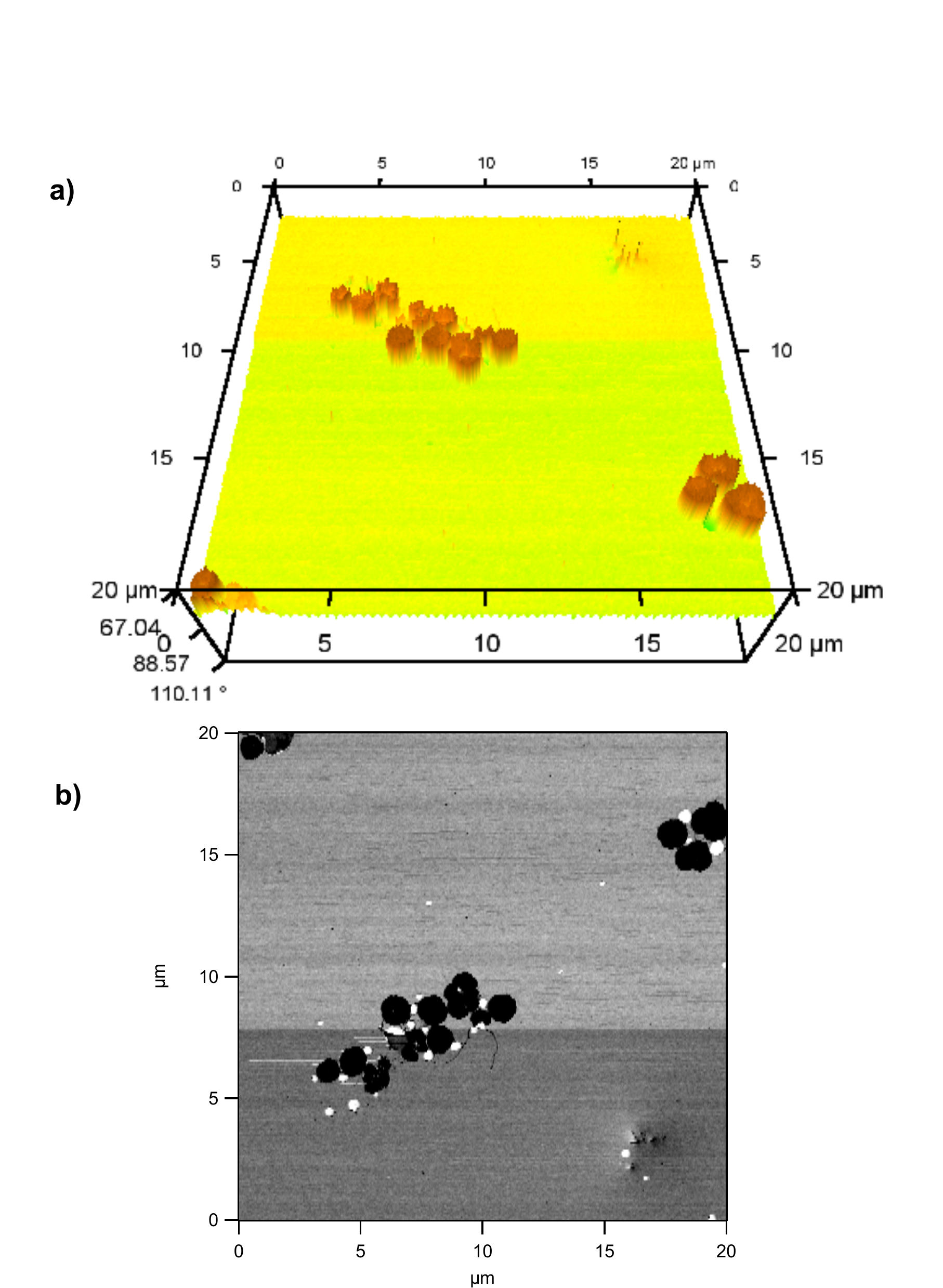}
\caption{ Irradiated sample with $30\, \text{Gy}$, a) 3D view of the AFM to 20$\, \mu m$, b) 2D view of the AFM to 20$\,\mu m$.}
\label{f:30gy2}
\end{figure}

After irradiation of the DNA which is submerged in water in a AFM, it is not possible to analyze the results through quantifying SSB and DSB damage and it was only possible to have a visualization from the changes. Thus, it was possible to observe the state of the DNA after interactions and when X-rays have left the DNA. Hence, I use fragment size of different applied doses for comparison.
For measuring the fragment size, I employ the Software Asylum Research. It uses the visualization images and improves their contrasts for better observation \cite{morris_atomic_1999}. Fig. \ref{f:AFM} shows the fragment size for different doses. It can be seen that the fragment sizes decrease with increasing the applied radiations. Reducing the size of fragments represents increasing the number of damage. Figures \ref{f:control} to \ref{f:30gy2} show the obtained images in the AFM. They confirm the reported results for fragment sizes. The analysis has been done based on standardized parameters of image analysis for the AFM.\\ \\
Moreover, I use a negative control sample to discover the effect of radiations on the DNA state in other samples.
Fig. \ref{f:control} shows the fragment size and DNA physical conditions of the negative control sample. It is observed that the DNA is a long fragment and consecutive. According to the indicated dimensions and conditions, it is possible to conclude this is for a double strand DNA chain \cite{eaton_atomic_2010}. I use this sample as negative control to compare with other samples which are under different radiation doses. Fig. \ref{f:0006gy} shows the fragmentation produced by a radiation equal to $0.006 \, \text{Gy}$. This value of radiation is achieved through using a constant energy value equal to $17.6 \,\text{KeV}$ when the exposition time is 2 seconds. Thus, through comparison between Fig. \ref{f:0006gy} and Fig. \ref{f:control}, it is possible to approximate the damage intensity based to the average fragment size.
Fig. \ref{f:0025gy} is corresponding to a radiation equal to $0.025 \, \text{Gy}$. This value of radiation is achieved through using a constant energy value equal to $17.6 \, \text{KeV}$ when the exposition time is $8$ seconds. It is possible to observe the formation of a higher DNA accumulation in some zones due to a large amount of particles which generate hits in these zones. They are contrasted with the zones where there is a lower accumulation and also with the negative control sample. Of the same way, the DNA in this image stops being consecutive and it is organised as expected in the treatments.\\ \\
Fig. \ref{f:008gy1} shows the fragmentation for the case with $0.08 \, \text{Gy}$ radiation. This value of radiation is achieved when the exposition time is $27$ seconds. The obtained results are similar to the $0.025 \,\text{Gy}$ case and the size and visualization of fragments in these cases are similar; however, there is no accumulation in the $0.08 \, \text{Gy}$ case. Fig. \ref{f:1gy1} is corresponding to a radiation equal to $1 \,\text{Gy}$. This value of radiation is achieved when the exposition time is 5 minutes and 35 seconds. This figure shows the DNA generating aggregations and packaging. It was observed the appearance of circles with a thickness equal to $5\,\mu \text{m}$. They can be the residues of some reactive parts in the preparation process of the sample. I improved the contrast of this image, and it was possible to observe these circles were coordinated with the DNA. Thus, it was possible to conclude that they are made of the same material. This conclusion is confirmed by \cite{cerreta_fine_2013}.\\ \\
Obtaining Fig. \ref{f:5gy1} and Fig. \ref{f:30gy2} was difficult to find in the AFM in comparison to the cases with lower radiations when it was consider the same conditions and DNA concentrations for all cases. The reason of this difficulty is that there was not enough DNA at the moment of mounting the sample over the MICA or the radiation is high for the sample's size and it causes sample's evaporation. This assumption was confirmed experimentally.
Fig. \ref{f:5gy1} is corresponding to a radiation equal to $5 \,\text{Gy}$. This value of radiation was achieved when the exposition time is $27$ minutes and 58 seconds. It was not possible to observe thinner helices in comparison to lower radiations. Furthermore, it was observed a large amount of thick circles which are coordinated with the DNA. Fig. \ref{f:30gy2} is corresponding to a radiation equal to $30\, \text{Gy}$. This sample is the
hardest case to detect and localize in the AFM.
The sample is observed in a sampling size equal to $20$$\,\mu \text{m}$. From a 3D view, the DNA is hard to distinguish; however, it is observed that the circles are not as uniform as the lower radiation cases. In the 2D view, it was possible to observe the DNA and many circles in the sample. The gray circles have the same DNA composition. On the other hand, the white circles have not the same DNA composition.

\section{Discussion}

Nowadays the real energy of the photons which are used for radiotherapy are in the order of a few $\text{MeV}$, however, the results presented in this paper consider energy values in the KeV regime given that comparison with the experimental results was desirable. Hence, the Compton Effect and the pair production effect are not predominant effects for the interactions between the rays and the DNA. For the KeV regime only the photoelectric effect is considered. For radiotherapy energies, photons should be as focused as possible to reduce risks \cite{elshaikh_advances_2006}. This energy could not be reach in the experimental setups mainly because the used PHYWE instrument do not have this value of energy. This energy value could only be reached with an exposition time in the order of days non-stopping.
For the simulation, it is necessary to increase the number of bases in the order of $10^{8}$ which is corresponding to the total length of the DNA. However, in the simulations due to computational limitations, it was only possible to run the simulation with $10000$ base pairs. It is worth mentioning that the program can hold any sequence length. Moreover, it is necessary to repeat the experimental part in the AFM since the protocol can have many errors and the damage analysis is done only according to the fragment size. Hence, it is ideal to increase the number of repetitions to have more samples, and thus, reaching a standard error lower than the data average.\\ \\
At the AFM sampling stage, it is possible to choose different zones of every sample since only a small zone of each sample is considered in every round of analysis. According to the simulation and experimental results, it is possible to conclude that there is a higher number of interactions for higher radiation doses. Moreover, the probability of DSB damage is increased with increasing the radiation doses. According to the simulation results, it is obtained that the maximum value of radiation without visible damage is equal to $0.025 \, \text{Gy}$, while according to the experimental results this value is lower and it is equal to $0.006 \, \text{Gy}$.\\ \\
It is possible to conclude that there is a probability to generate damage in DNA even for radiations with low doses and damage can be lethal or not. By repeating simulation, it is possible to observe that at least one event occurs at low doses and there is no restriction for this event to happen in both simulation and real scenarios. In addition, it is possible to simulate superior levels of organisation of the DNA by considerations of chromatin fibers packing. However, due to the time limitation, it was not possible to reach this characterisation. It is worth mentioning that it is possible to create superior levels of DNA organization using the same geometrical principle of the first structural level. Superior packaging simulations can show the important factors for DNA protection and enhancing the resistance of DNA to the damage \cite{friedland_stochastic_2010}.

\section{Conclusions and Perspectives}
This simulation is a significant advancement for DNA modelling which takes into account the atomic composition of DNA and its sequence and considers them as crucial parameters in the damage generated through interaction with X-rays. However, this simulation requires a large amount of computational power if it is necessary to obtain results in larger scale scenarios.
The results that have been found are confirmed by the reported results in the literature. When the value of radiation is increased the number of events which contain the SSB and DSB damage also increases. For the energy considered here, the effect is linear and with different slopes depending on the radiation values. The perspective for the future of this research is to achieve a software which could be freely implemented in radiotherapy treatments. The aim of this software is to predict in a more accurate way, the DNA damage after radiotherapy. Hence, it is possible to estimate the exposition time with X-rays and the radiation doses that can be received by patients without generation of higher damage. Adding the DNA reparation process to the employed model is difficult since it is specific for each patient and it depends on the mass of the irradiated zone.

\section{Acknowledgements}
I would like to give special acknowledgement to Keyvan Aghababaiyan, who motivated me to publish the results of this research and who helped me translating it and editing it. Special gratitude to Professor Helena Groot, who guided me and advised me during the experimental development of this research. Thank you so much to Los Andes University and the Human Genetics Laboratory for their complete support and for providing the instruments and resources for the development of this investigation. Special gratitude to the High Energy Physics Laboratory and their support with the Atomic Force Microscopy. This research was not founded by any institution.
\bibliography{ref2}

\end{document}